\documentclass[twocolumn]{aastex61}

\usepackage{amsmath,tabu}
\usepackage{booktabs}

\shorttitle{X-ray induced deuterium enrichment of organics in protoplanetary disks}
\shortauthors{Gavilan et al.}

\begin{document}

\title{X-ray induced Deuterium Enrichment of N-rich organics in protoplanetary disks: \\  an experimental  investigation using synchrotron light}  

\correspondingauthor{Lisseth Gavilan}
\email{lisseth.gavilan@latmos.ipsl.fr}

\author[0000-0002-0786-7307]{Lisseth Gavilan}
\affil{LATMOS, Universit\'e Versailles St Quentin, UPMC Universit\'e Paris 06, CNRS, 11 blvd d'Alembert, 78280 Guyancourt, France}

\author{Laurent Remusat}
\affiliation{IMPMC, CNRS UMR 7590; Sorbonne Universit\'es, UPMC Universit\'e Paris 06;  IRD, Mus\'eum National d'Histoire Naturelle, CP 52, 57 rue Cuvier, Paris 75231, France}

\author{Mathieu Roskosz}
\affiliation{IMPMC, CNRS UMR 7590; Sorbonne Universit\'es, UPMC Universit\'e Paris 06;  IRD, Mus\'eum National d'Histoire Naturelle, CP 52, 57 rue Cuvier, Paris 75231, France}

\author{Horia Popescu}
\affiliation{SEXTANTS beamline, SOLEIL synchrotron, L'Orme des Merisiers, 91190 Saint-Aubin, France}

\author{Nicolas Jaouen}
\affiliation{SEXTANTS beamline, SOLEIL synchrotron, L'Orme des Merisiers, 91190 Saint-Aubin, France}

\author{Christophe Sandt}
\affiliation{SMIS beamline, SOLEIL synchrotron, L'Orme des Merisiers, 91190 Saint-Aubin, France}

\author{Cornelia J\"ager}
\affiliation{Laboratory Astrophysics and Cluster Physics Group of the Max Planck Institute for Astronomy at the Friedrich Schiller University \& Institute of Solid State Physics, Helmholtzweg 3, 07743 Jena, Germany}

\author{Thomas Henning}
\affiliation{Max-Planck Institute for Astronomy K\"onigstuhl 17, 69117 Heidelberg, Germany}

\author{Alexandre Simionovici}
\affiliation{Institut des Sciences de la Terre, Observatoire des Sciences de l'Univers de Grenoble, BP 53, 38041 Grenoble, France}

\author{Jean Louis Lemaire}
\affiliation{Institut des Sciences Mol\'eculaires d'Orsay (ISMO), CNRS, Univ. Paris Sud, Universit\'e Paris-Saclay, 91405 Orsay, France}

\author{Denis Mangin}
\affiliation{Institut Jean Lamour, CNRS, Universit\'e de Lorraine, 54011 Nancy, France }

\author{Nathalie Carrasco}
\affil{LATMOS, Universit\'e Versailles St Quentin, UPMC Universit\'e Paris 06, CNRS, 11 blvd d'Alembert, 78280 Guyancourt, France}

\begin{abstract}

The deuterium enrichment of organics in the interstellar medium, protoplanetary disks and meteorites has been proposed to be the result of ionizing radiation. The goal of this study is to simulate and quantify the effects of soft X-rays (0.1 - 2 keV), an important component of stellar radiation fields illuminating protoplanetary disks, on the refractory organics present in the disks. We prepared tholins, nitrogen-rich organic analogs to solids found in several astrophysical environments, e.g. Titan's atmosphere, cometary surfaces and protoplanetary disks, via plasma deposition. Controlled irradiation experiments with soft X-rays at 0.5 and 1.3 keV were performed at the SEXTANTS beam line of the SOLEIL synchrotron, and were immediately followed by ex-situ infrared, Raman and isotopic diagnostics. Infrared spectroscopy revealed the preferential loss of singly-bonded groups (N-H, C-H and R-N$\equiv$C) and the formation of  sp$^3$ carbon defects with signatures at $\sim$1250 - 1300 cm$^{-1}$. Raman analysis revealed that, while the length of polyaromatic units is only slightly modified, the introduction of defects leads to structural amorphization. Finally, tholins were measured via secondary ion mass spectrometry (SIMS) to quantify the D, H and C elemental abundances in the irradiated versus non-irradiated areas. Isotopic analysis revealed that significant D-enrichment is induced by X-ray irradiation. Our results are compared to previous experimental studies involving the thermal degradation and electron irradiation of organics. The penetration depth of soft X-rays in $\mu$m-sized tholins leads to volume rather than surface modifications:  lower energy X-rays (0.5 keV) induce a larger D-enrichment than 1.3 keV X-rays, reaching a plateau for doses larger than 5 $\times$ 10$^{27}$ eV cm$^{-3}$. Synchrotron fluences fall within the expected soft X-ray fluences in protoplanetary disks, and thus provide evidence of a new non-thermal pathway to deuterium fractionation of organic matter.

\end{abstract}


\section{Introduction}
\label{sec:intro}

The study of organic molecules, dust and ices is essential to understand the origin and evolution of the solar system. The deuterium to hydrogen ratio (D/H) is used as a proxy of the surrounding physico-chemical conditions affecting organics, omnipresent in inter- and circumstellar media (ISM/CSM), and occasionally delivered to Earth and other planets in the form of meteorites.  \cite{Kerridge1983} proposed early on that  deuterium enrichment in meteorites had an interstellar origin. Yet interstellar clouds can further evolve, reaching a critical mass and density, collapsing, and forming protoplanetary disks \citep{Williams2011} on par with our own protosolar nebula (PSN). Isotopic fractionation can occur at any of these stages via multiple chemical pathways such as isotopic exchange reactions in the ISM \citep{Roueff2015} and gas-surface reactions in protoplanetary disks \citep{Henning2013}. 

Deuterated molecules have been detected in all stages of star formation; for a review, see \cite{Ceccarelli2014}. In molecular clouds, detected deuterated species include DCO$^+$ and DNC \citep{VanderTak2009}, H$_2$D$^+$ and HD$_2^+$  \citep{Parise2011} among many others. In proto-stellar cores, detections include DCN and D$_2$CO \citep{Roberts2002, Ceccarelli2002}, CH$_3$CCD \citep{Markwick2005}, CH$_3$OD \citep{Parise2006}, NH$_2$CDO and DNCO \citep{Coutens2016}. In protoplanetary disks such as TW Hydrae, the most well-studied protoplanetary disk because of its proximity and near face-on orientation, detections include DCO$^+$  \citep{VanDishoeck2003, Qi2008}, H$_2$D$^+$ \citep{Ceccarelli2004} and DCN \citep{Qi2008, Oberg2012}. In the young protoplanetary disk surrounding the solar-type protostar DM Tau, detections of H$_2$D$^+$ \citep{Ceccarelli2004} and HDO \citep{Ceccarelli2005} were reported but later contested by \cite{Guilloteau2006}, along with the confirmed detection of DCO$^+$. Chemical models including a vast network of deuterated species have pointed at the PSN origin of water found in comets and asteroids \citep{Albertsson2014}, although modeling work by \citep{Cleeves2014} argued that ion-driven deuterium pathways in the disk are inefficient, implying an interstellar origin for water. Therefore, the dominance of deuterium fractionation at a particular stage in stellar evolution persists as an open question in astrophysics. 

In the solar system, deuterated species have been measured in comets and meteorites. 
Recent in situ measurements in the Jupiter family comet 67P/Churyumov-Gerasimenko by the \textsc{Rosetta} spacecraft found the D/H in water to be approximately three times the terrestrial value \citep{Altwegg2015}, suggesting an asteroidal rather than cometary origin for the Earth oceans and atmosphere.  This contrasts with reports on the D/H ratio of the Jupiter-family comet 103P/Hartley 2 \citep{Hartogh2011}, compatible with the value of Earth ocean-like water. These measurements point at mixing processes in the early PSN. Analysis of meteorites have shown deuterium enrichment of their insoluble organic matter (IOM) compared to the solar hydrogen abundance \citep{Robert1982, Yang1983}, and suggested an interstellar inheritance \citep{Sandford2001, Busemann2006}. D-enrichment has also been measured in interplanetary dust particles (IDPs) \citep{Messenger2000} and more recently in UltraCarbonaceous Antarctic Micrometeorites (UCAMMs) \citep{Duprat2010}. However, isotopic anomalies in meteorites and comets are not necessarily fingerprints of an interstellar origin: organic matter can be further modified in the coldest outer regions of protoplanetary disks  \citep{Remusat2006, Aleon2010}, e.g. via isotopic gas-grain exchange reactions  \citep{Remusat2009}. 

Laboratory experiments have the potential of unveiling the chemical and physical mechanisms at the origin of isotopic fractionation. Thermal processing experiments can impact the chemical and structural properties of astrophysical dust and ice analogs, e.g. \citep{Allamandola1988, Ehrenfreund1999, Collings2004}. However, non-thermal irradiation processes are as important in astrophysical environments where elevated dust or gas densities inhibit thermal annealing. Ionizing radiation includes charged particles (electrons or protons), ions, but also UV photons and X-rays \citep{Feigelson2002}. Recent experiments have investigated the effects of electron irradiation on the structure and the D/H signature of kerogens (natural organic macromolecules) \citep{LeGuillou2013}, synthetic polymeric organics \citep{Laurent2014, Laurent2015} and in analogs of protoplanetary hydrous silicate dust \citep{Roskosz2016}. These experiments have revealed that non-thermal irradiation can offer viable pathways to deuterium enrichment in the solid phase.

X-rays from T-Tauri stars are associated to the energy dissipation of magnetic coronal fields \citep{Preibisch2005}. Together with UV photons, these can irradiate the disk surface, mid-plane and outer edges promoting photochemistry and radiolysis \citep{Glassgold1997, Henning2010, Aresu2011}. As of now, no laboratory investigations have considered the possible isotopic effects of X-ray irradiation on organic matter.  Yet, we know hard X-rays ($>$ 5 keV) can strongly impact the structure of silicates in protoplanetary disks \citep{Gavilan2016b}. Thus, our goal is to experimentally unveil the isotopic effects of stellar soft X-rays (0.1 - 2 keV) on protoplanetary organic matter. 

To this end, we employ nitrogen-rich amorphous organic solids as analogs to refractory organics produced from the thermal and/or photo-processing of N-rich ices in regions beyond the snow-line of protoplanetary disks \citep{Greenberg1995, Ciesla2012}. To simulate the T-Tauri X-ray field, we perform controlled experiments using synchrotron radiation in the soft X-ray range. The brilliance of synchrotron X-rays is ideal to perform highly-focused monochromatic studies with remarkable reliability, allowing us to simulate astrophysical fluences. 

\section{Experiments}

\subsection{Sample preparation}
\label{sec:ellipso}
\begin{figure}[ht!]
\gridline{\fig{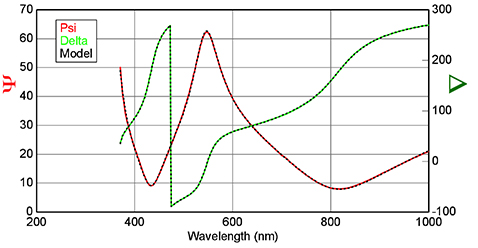}{0.42\textwidth}{(a)}  }
\gridline{\fig{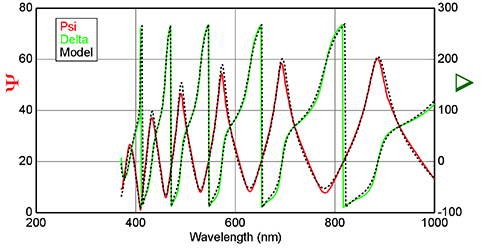}{0.42\textwidth}{(b)}  }
\caption{Ellipsometric parameters $\Psi$ and $\Delta$ measured in the UV-Visible (390 - 1000 nm) and modeled to determine the thickness of \textbf{(a)} a thin tholin film, and \textbf{(b)} a thick tholin film, used for X-ray absorption spectroscopy and irradiation. \label{fig:psi}}
\end{figure}

Tholins, complex organic compounds rich in carbon and nitrogen used as analogs to astrophysical solid matter \citep{Sagan1979}, were produced using the \mbox{PAMPRE} setup, a low pressure (0.95 mbar) radiofrequency (RF) plasma reactor located at LATMOS (Guyancourt, France) \citep{Szopa2006}. A 13.56 MHz RF power source tuned at 30 W generates a capacitively coupled plasma (CCP) fed by a gas mixture of N$_2$:CH$_4$ = 95:5, conditions chosen for optimal tholin production \citep{Sciamma2012}. The plasma was produced within a vertical cylindrical cage and the substrates were placed on the bottom electrode. The produced films, rich in carbon and nitrogen, i.e. N/C = 0.2, H/C $\sim$0.1 \citep{Carrasco2016},  make them comparable to the organic matter in UCCAMs with N/C = 0.05 - 0.12 \citep{Dartois2013} and to refractory organics that may have formed from the energetic irradiation of nitrogen-rich ices in the outer solar system.

Two types of tholins with different thicknesses were prepared. A thin film ($\sim$300 nm) on an X-ray transparent substrate was prepared for X-ray absorption spectroscopy with the goal of obtaining absolute absorption coefficients in the soft X-ray range. Thick films ($\sim$1200 nm) were prepared for irradiation with the goal of absorbing most of the deposited X-rays within the sample volume. The thin tholin was produced after turning on the plasma for 30 minutes. It is evenly deposited on both the Si$_3$N$_4$ membrane and the supporting frame made of silicon (5 $\times$ 5 mm). For this sample, ellipsometry is performed on the reflective Si frame. The Si$_3$N$_4$ itself is 30 nm thick and quasi-transparent to X-rays \citep{Torma2014}.  
Thick tholin films were deposited on single-side polished silicon substrates (9 $\times$ 4 mm) after leaving the plasma on for 150 minutes.

To determine the film thickness, we used spectroscopic ellipsometry (SE). This is an optical technique where a beam of polarized light is obliquely sent towards a film deposited on a reflective surface. Light is elliptically polarized upon reflection. Both $tan(\Psi)$, the ratio of the incident to the reflected light amplitude, and $\Delta$, the relative phase change for \textit{p}- and \textit{s}-polarized light components, are measured, as seen in Fig. \ref{fig:psi}.
These determine the absolute value of the complex reflectance ratio, $\rho$, defined as,
\begin{equation}
\rho = \dfrac{r_p}{r_s}=tan(\Psi)e^{(i\Delta)}
\end{equation}
where $r_P$ and $r_S$ are the complex Fresnel reflection coefficients of \textit{p}- and \textit{s}-polarized light. Using $\rho$ and an appropriate optical model we can find the $\chi$-minimized dielectric properties, layer (or multi-layer) film thickness, and roughness. For this study we used a rotating spectroscopic ellipsometer (M-2000V, \textit{J.A. Woollam}), which allows all wavelengths to be measured simultaneously. The light source is a tungsten lamp covering the 370 - 1000 nm range, incident at an angle of 70$^{\circ}$ and with a spot size at the sample of $\sim$2 $\times$ 3 mm. Modeling of the ellipsometric parameters was done with the   Complete-EASE$^{TM}$ ellipsometry software. We selected a Tauc-Lorentz model, suitable for amorphous organic films \citep{Jellison1996}, providing the thickness $d$, surface roughness $r$, E$_g$ (the Tauc optical gap), and the mean squared error (MSE) between the modeled and the experimental data.  The modeled parameters are listed in Table \ref{tab:ellip}, showing the consistent gap value (and thus homogeneity) of films prepared under the same conditions but at different durations, in agreement with previous measurements \citep{Mahjoub2012}.

\begin{table}[htbp!]
\footnotesize
\centering
\caption{Ellipsometric parameters for two tholin films used for XAS and irradiation respectively.}
\label{tab:ellip}
\begin{tabular}{cccc}
\toprule
MSE        & Thickness [ nm ] & Roughness [ nm ] & E$_g$ [ eV ]    \\
\hline
3.2$\times$10$^{-3}$ & 305  $\pm$ 12     & 5.1 $\pm$ 1.1     & 2.12 $\pm$ 0.08 \\
8.2$\times$10$^{-3}$ & 1139 $\pm$ 42     & 13.2 $\pm$ 2.1      & 2.10 $\pm$ 0.05\\
\bottomrule
\end{tabular}
\end{table}
 
\subsubsection{X-ray absorption spectroscopy}

\begin{figure}[h!]
\centering
\epsscale{1.15}
\plotone{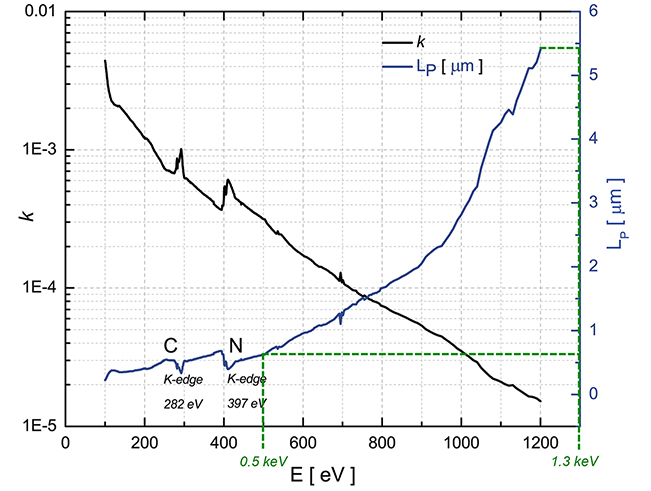}
\caption{Imaginary index of refraction as a function of energy and calculated penetration length, L$_P$, in $\mu$m. The energies selected for irradiation are marked in green. \label{fig:klp}}
\end{figure}

We performed X-ray absorption spectroscopy (XAS) measurements at the COMET experimental station, an ultra-high vacuum chamber placed in the intermediate focal point of the elastic branch of the SEXTANTS beamline \citep{Sacchi2013} of the SOLEIL synchrotron.  The samples were placed on a motorized sample holder perpendicular to the X-ray beam. Four different gratings were used: 100 - 300 eV, 300 - 500 eV, 500 - 800 eV and 800 - 1200 eV with resolution E/$\Delta$E $>$10$^4$. For each range two samples were measured: the thin tholin on a Si$_3$N$_4$ membrane and a bare Si$_3$N$_4$ membrane. The  optical depth is obtained using the Beer-Lambert law,
\begin{equation}
\tau(\nu) =-\ln\dfrac{I(\nu)}{I_0(\nu)},
\end{equation}
where $I(\nu)$ is the intensity of light at a given frequency, $\nu$, transmitted through the tholin film deposited on the membrane and $I_0$$(\nu)$ is the light intensity through the membrane alone. The imaginary index of refraction can be obtained via,

\begin{equation}
k = \dfrac{\tau(\nu)}{4 \pi \nu d}
\end{equation}

where $d$, the film thickness, is found by ellipsometry as aforementioned. Similarly, the penetration depth ($L_P$) of a photon is given by,

\begin{equation}
L_P ( \mu m ) = \dfrac{\lambda (\mu m) }{4 \pi k}
\end{equation}

The resulting $k$ and $L_P$ ($\mu$m) in the soft X-ray range for the tholins is presented in Fig. \ref{fig:klp}. Fluctuations in the $k$ value around 100 eV are due to Si L edges from the Si$_3$N$_4$ substrate. In addition to the expected C K-edge (390 eV) and N K-edge (397 eV) absorptions, we also notice the O K-edge (530 eV) and the Fe L-edges (720 eV). These are attributed to pollution and sputtering of the confining plasma cage where tholins are produced.  
Both $k$ and $L_P$ are the required parameters to optimize the irradiation experiments on thick tholins. While 0.5 keV X-rays penetrate the tholin film up to a thickness of $\sim$625 nm (depositing most of their energy within the 1.16 $\mu$m film), 1.3 keV X-rays have a penetration depth greater than 5 $\mu$m (i.e. only a fraction of their energy will be deposited within the film, the other will be scattered). In this paper, we report the incident dose on each experiment and not the adsorbed dose, which varies depending on the chemical nature of the material (reflected in its absorption index) and on its thickness.

\subsection{X-ray irradiation}
\begin{figure}[htbp!]
\epsscale{1.15}
\plotone{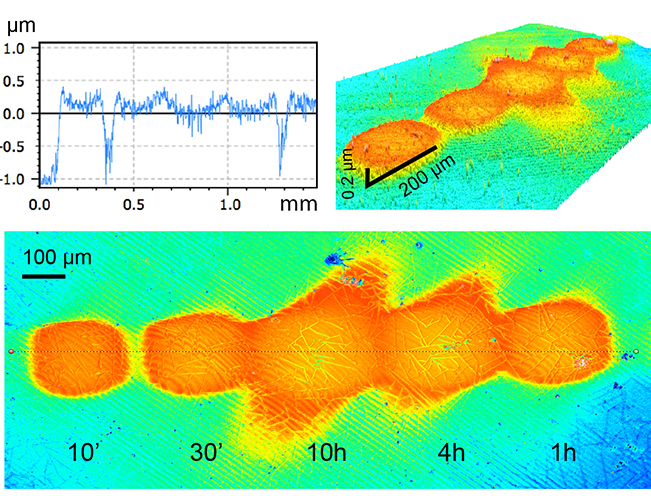}
\caption{\textit{Top}: 2D and 3D topographies measured via optical interferometry on the 0.5 keV irradiated tholin showing crater-like features. \textit{Bottom}: 2D image of irradiated spots following X-ray doses at 0.5 keV.\label{fig:profilo}}
\end{figure}
We performed the X-ray irradiation experiments at the COMET station using two monochromatic settings, chosen to evaluate the impact of irradiation across the soft X-ray region: 0.5 keV and 1.3 keV. The beam was right-circularly polarized for both experiments with a spot size at the sample of 80 $\times$ 50 $\mu$m, flux density $\sim$10$^{17}$ photons s$^{-1}$ cm$^{-2}$ and resolution of E/$\Delta$E $>$ 10$^4$ up to 1200 eV. For each energy, tholin films were irradiated at spots separated by 300 $\mu$m for different durations: 10 minutes, 30 minutes, 1 hour, 4 hours and 10 hours.  

Following irradiation, the sample surface was examined with a white-light interferometer (\textit{smartWLI}), which allowed us to record 3-D topographies with depth resolution $\sim$3 nm. The result is shown in Fig. \ref{fig:profilo}, where X-irradiated regions are manifested by  volume expansion seen as elevated craters on the tholin surface (with rising edges up to 200 nm from the surface), reticulation and darkening. 
The irradiated spots extend up to $\sim$200 $\times$ 200 $\mu$m$^2$, i.e. twice as large as the incident beam size, revealing the diffusion of secondary electrons during irradiation.  The central spot, corresponding to the highest X-ray dose, extends even farther out than all others.  
  
\section{Non-destructive diagnostics}
We performed ex-situ diagnostics of the irradiated tholins via infrared and Raman microspectrometry on the 0.5 keV irradiated tholin. These diagnostics allow us to further quantify the chemical and/or structural evolution of the irradiated tholin films, revealed by optical interferometry in the previous section. Access to infrared and Raman microscopes was kindly provided by the SMIS beamline \citep{Dumas2006} at SOLEIL.

\subsection{Infrared Microscopy}
\label{sec:ir} 
 
\begin{figure}[htbp!]
\epsscale{1.15}
\plotone{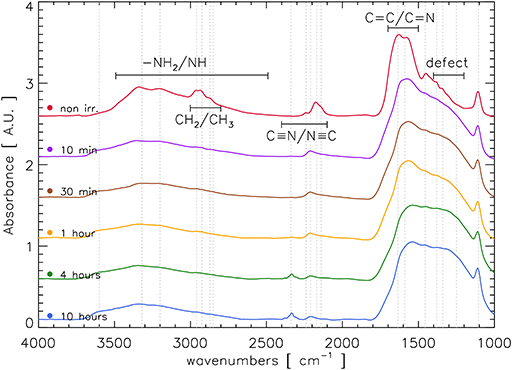}
\caption{Evolution of infrared spectra of irradiated tholins at increasing X-ray doses at 0.5 keV. The peak at 1100 cm$^{-1}$ is attributed to the Si substrate. \label{fig:ir1}}
\end{figure}
 
Infrared spectra of irradiated and non-irradiated regions were measured to quantify the chemical evolution of the irradiated films at increasing X-ray doses. We used a iN10 microscope (\textit{ThermoFisher Scientific}) coupled to a FTIR spectrometer in offline mode. Areas of 100 $\times$ 100 $\mu$$m^2$ within the irradiated regions were integrated from 4000 to 1000 cm$^{-1}$. A polynomial baseline was subtracted from the infrared spectra to normalize the continuum arising from rough surface scattering and from broad ultraviolet absorptions typical of amorphous carbon materials reaching the near-infrared \citep{Khare1984, Bernard2006}.

\begin{figure}[htbp!]
\gridline{\fig{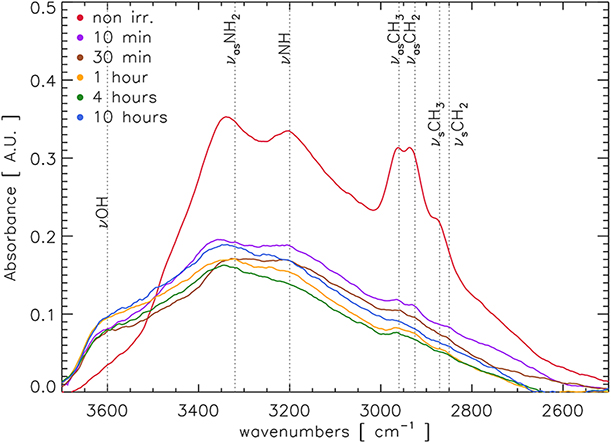}{0.34\textwidth}{(a)} }
\gridline{\fig{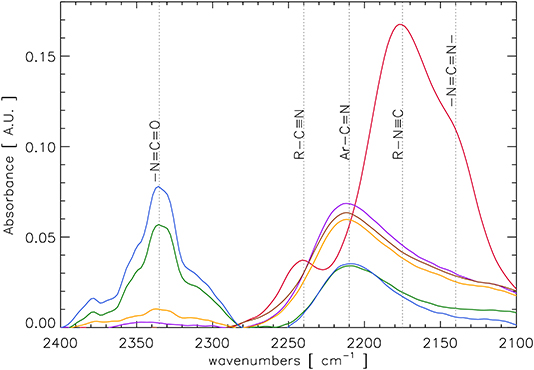}{0.355\textwidth}{(b)} } 
\gridline{\fig{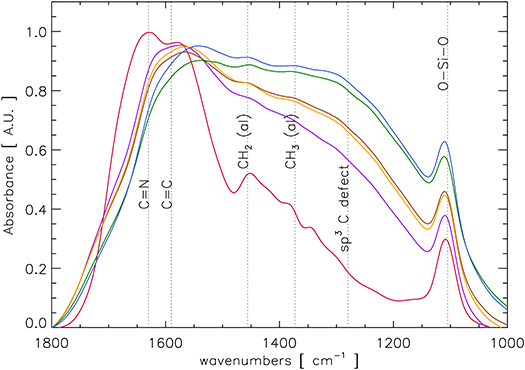}{0.35\textwidth}{(c)} }   
\caption{Infrared spectra of irradiated tholins at 0.5 keV split into main spectral regions as defined in Section \ref{sec:ir}. \label{fig:ir2}}
\end{figure}

\begin{figure}[h!]
\gridline{\fig{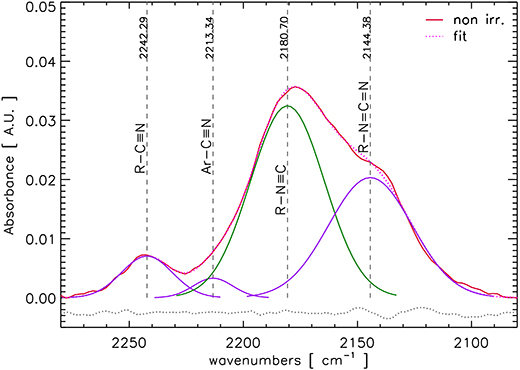}{0.35\textwidth}{(a)}}
\gridline{\fig{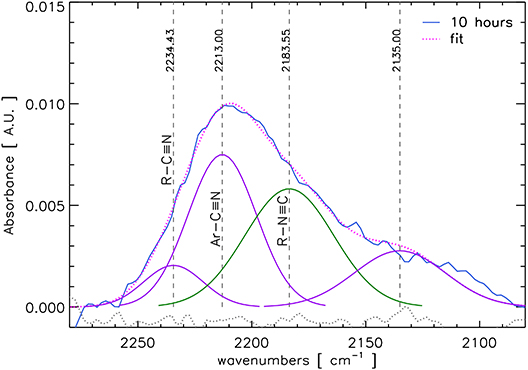}{0.35\textwidth}{(b)}}
\caption{Infrared band evolution of the nitrile bands for X-ray doses at 0.5 keV. Deconvolution of the main nitrile modes for \textbf{(a)} the initial tholin, \textbf{(b)} the 10 hour irradiated tholin, showing the loss of the R-N$\equiv$C band intensity for the growth of the Ar-C$\equiv$N band intensity \label{fig:NC}}
\end{figure}

Fig.~\ref{fig:ir1} shows the evolution of the infrared vibrational bands from the initial (non-irradiated) tholin to the increasingly irradiated regions. This is detailed in Fig. \ref{fig:ir2}, where infrared spectra have been split into three main regions: (a) the 3600 - 2500 cm$^{-1}$ region, which includes stretching modes of amine groups (primary/secondary), superimposed by the methyl (CH$_3$) and methylene (CH$_2$) stretching modes, (b) the 2400 - 2000 cm$^{-1}$ region, which includes nitriles and isonitriles (C$\equiv$N/N$\equiv$C) stretching modes and (c) the 1800 to 1000 cm$^{-1}$ region which includes several overlapping bands including the sp$^2$ aromatic C=C and C=N stretching modes, sp$^3$-CH$_{2,3}$ bending modes, and sp$^3$-like C-C or C-N stretching of defects (such as non-hexagonal rings of aromatic compounds \citep{Carpentier2012}). For a complete attribution to vibrational modes of tholins in the mid-infrared we refer to \cite{Imanaka2004}. 
To analyze the kinetic evolution of the main vibrational modes, we performed the deconvolution of the continuum subtracted 3.4 $\mu$m band (3100 - 2800 cm$^{-1}$) and on the nitrile bands (2300 - 2000 cm$^{-1}$), presented in Fig. \ref{fig:NC}. Integrated optical depths were fitted as a function of the irradiated dose with the following exponential relation,
\begin{gather}
y = y_o + Ae^{-kE_{dose}},
\end{gather}
where $y$ corresponds to the integrated optical depths , y$_0$ corresponds to an asymptotic steady-state, $A$ is a constant, $k$ is the rate constant , and E$_{dose}$ is the irradiation dose.

\begin{figure}[htb!]
\epsscale{0.9}
\plotone{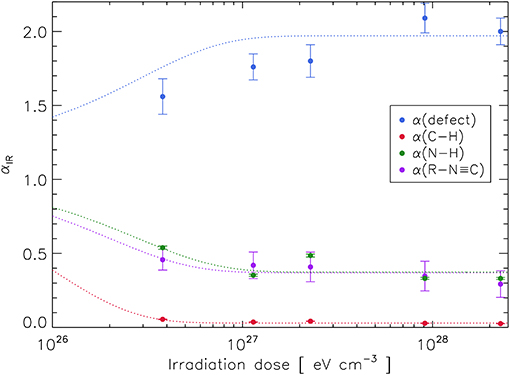}
\caption{Evolution of the intensities of the main infrared bands (normalized to the integrated optical depth of the non-irradiated tholin infrared bands) as a function of the deposited X-ray dose, displaying the favored depletion of alkyls, amines and singly bonded isonitriles, and the growth of sp$^3$ carbon defects. IR band kinetics are fitted with exponential functions. \label{fig:irkin}}
\end{figure}

\begin{figure}[htb!]
\epsscale{1.}
\plotone{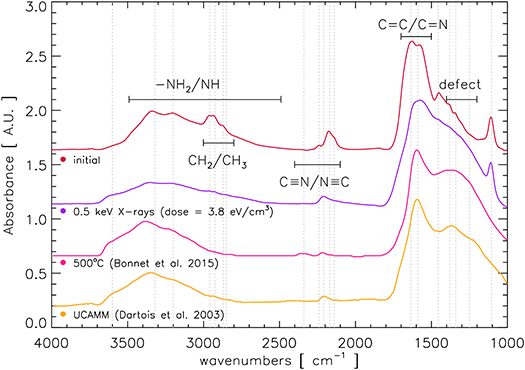}
\caption{\textit{From top to bottom}: Comparison between infrared spectra of an initial tholin to an X-irradiated tholin, thermally annealed tholin \citep{Bonnet2015}, and a UCAMM \citep{Dartois2013}.\label{fig:ir3}}
\end{figure}

Fig. \ref{fig:irkin} shows the infrared band kinetics during X-ray irradiation. The fitted parameters are reported in Table \ref{tab:fit}. We note the rapid depletion of the C-H stretching band (2800 - 3000 cm$^{-1}$) by the 95\% loss of its original intensity and the depletion of the N-H band and the R-N$\equiv$C at 2175 cm$^{-1}$ band both by the $\sim$67\% loss of their original intensity following the maximum X-ray doses. Before the continuum baseline correction, we note that the absorbance at 4000 cm$^{-1}$ increases from 0.7 to 0.9 at the highest X-ray dose, signaling darkening of the tholin film. 

We also note the growth of specific band intensities during irradiation, such as the vibrational mode at 2210 cm$^{-1}$, attributed to -CN groups attached to aromatic molecules \citep{Khare1981, Sagan1993},  and the sp$^3$ carbon defect band between 1250 - 1300 cm$^{-1}$, affecting the entire bending mode region between 1800 - 1000 cm$^{-1}$. Nitriles are not completely lost but the network bearing them becomes unsaturated and more aromatic. 
For the longest exposures (4 and 10 hours), a new peak appearing at 2335 cm$^{-1}$ is attributed to isocyanate R-N=C=O modes, likely the result of photo-oxidation of impurities in the tholin (noticed by the weak O edge in the X-ray absorption spectra, Fig. \ref{fig:klp}). In Fig \ref{fig:ir3}, we further compare the infrared spectra of our X-irradiated tholins to a thermally annealed tholin \citep{Bonnet2015} and to a UCAMM \citep{Dartois2013}. While the X-irradiated tholin does not present the high loss of amines as in the other two, the growth of the defect band near 1300 cm$^{-1}$ occurs in the three cases.  

\subsection{Raman microspectrometry}
\begin{figure}[ht!]
\epsscale{1.}
%\plotone{xraman.jpg}
\plotone{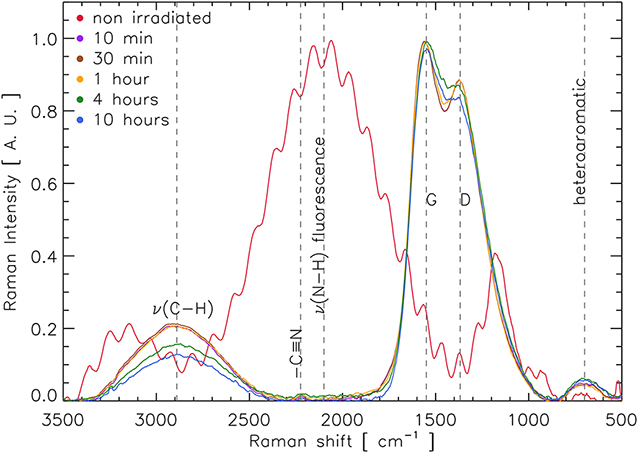}
\caption{Raman spectra of X-ray irradiated tholins  as a function of X-ray dose at 0.5 keV. The strong fluorescence from the non-irradiated tholin hinders measurement using a 514 nm excitation source. \label{fig:raman1}}
\end{figure}

Raman spectra of irradiated and non-irradiated regions were measured to quantify the structural evolution (size of aromatic units, graphitization/amorphization) of the irradiated films at increasing X-ray doses. These were measured with a DXR Raman spectrometer (\textit{ThermoFisher Scientific}) using a 514 nm excitation laser, with laser power $\le$ 0.3 mW focused with a 50$\times$ objective and a spot diameter of 2 - 3 $\mu$m, well within the 200 $\times$ 200 $\mu$m$^2$ irradiated regions (Fig. \ref{fig:profilo}). 

The main features in the Raman spectra of carbon-rich organic materials are the so-called G and D peaks, around $\omega_G$ = 1560 and $\omega_D$ = 1360 cm$^{-1}$ dominated by sp$^2$ sites, as visible excitation resonates with $\pi$ states. We note that the non-irradiated tholin produces an intense fluorescence background in the Raman signal, an issue that could be avoided via photobleaching (which inherently transforms the material) or by employing an ultraviolet excitation source \citep{Bernard2006}. This was not a problem for the X-irradiated, nitrogen-depleted, regions. The baseline corrected Raman spectra of the irradiated tholins are shown in Fig. \ref{fig:raman1}, normalized to the intensity of the G band. For the non-irradiated tholin, measurement of the Raman spectra was hindered by strong fluorescence. 
Raman spectra were deconvolved following the prescription in \cite{Ferrari2000}, where the G band is fitted with a Breit-Wigner-Fano profile and the D band is fitted with a Lorentzian profile. This procedure provides Raman parameters including the position of each band, their full-width at half maximum (FWHM), and the ratio of their integrated intensities (I$_D$/I$_G$). These parameters are listed in Table \ref{tab:raman}. 

\begin{figure}[htbp!]
\gridline{\fig{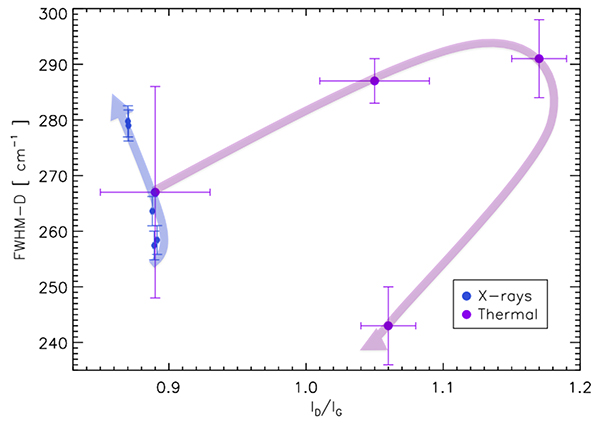}{0.4\textwidth}{(a)} }
\gridline{ \fig{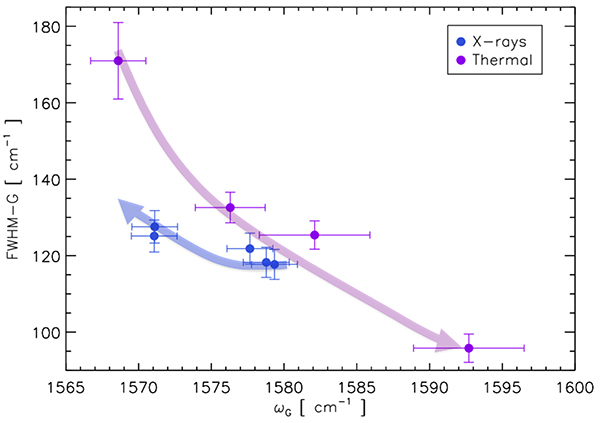}{0.4\textwidth}{(b)}          }
\caption{Raman parameters of irradiated tholins as a function of the X-ray dose at 0.5 keV compared to the evolution of thermally annealed (300 - 1000 $^{\circ}$C) tholins \citep{Bonnet2015}. \textbf{(a)} The size of polyaromatic units ($\propto$ I$_D$/I$_G$) is slightly affected by X-ray irradiation, while strongly affected by thermal heating. \textbf{(b)} Defect creation leads to amorphization in X-ray irradiated tholins, while aromatization dominates over defect creation in thermally degraded tholins. \label{fig:raman3}}
\end{figure}

As proposed in \cite{Ferrari2007}, the Raman parameters for completely disordered, amorphous carbons consisting of distorted sixfold rings are related as follows,
\begin{equation}
I_D/I_G = C' L_a^2,
\end{equation}
where C'(514 nm) = 0.0055, and $L_a$ is the average correlation length of polyaromatic units in \AA. For the X-ray irradiated tholins, the size of polyaromatic units remains almost unchanged, i.e. $L_a$ increases very slightly after 30 minutes irradiation and barely decreases from 1.27 $\pm$0.02 to 1.25 $\pm$0.02 nm after 10 hours, implying that no larger polyaromatic structures are created. As the X-ray irradiation dose increases, the intensity of the D band decreases with respect to the G band, while both the FWHM-G and FWHM-D increase. The broadening and position shift are attributed to an increasing number of defects, including bond length and angle disorder at the atomic scale resulting from cross-linking due to new bond formation. This leads to the reorganization in the tholin structure manifested macroscopically as reticulation and volume expansion (Fig. \ref{fig:profilo}). 

In Fig. \ref{fig:raman3}, we compare the evolution of Raman parameters of tholins undergoing X-ray irradiation (this study) to tholins that are thermally degraded \citep{Bonnet2015}. For the latter, heating leads not only to carbonization (preferential expulsion of heteroatoms) but also to the creation of polyaromatic structures, while in our case polyaromatic structures are almost unaffected by X-ray irradiation. However, X-ray irradiation may favor the creation of isolated over polyaromatic units, as signaled from the infrared analysis of the nitrile modes in Section \ref{sec:ir}.

\section{Destructive diagnostics: SIMS}

\begin{figure}[h!]
\epsscale{1.15}
\plotone{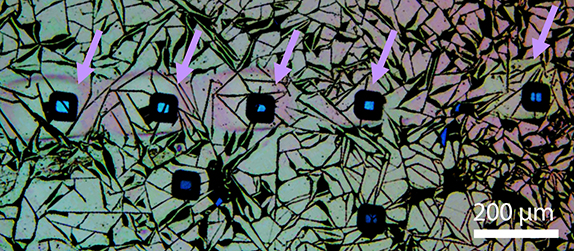}
\caption{X-ray (1.3 keV) irradiated tholin following SIMS analysis of irradiated regions (brighter spots indicated by arrows) and the reference (non-irradiated) regions. Sputtering stops as soon as the Si substrate is reached (blue spots). \label{fig:simsbino}}
\end{figure}

We used Secondary Ion Mass Spectroscopy (SIMS) to characterize the evolution of elemental abundances of the irradiated tholins at 0.5 keV and 1.3 keV and to detect possible isotopic effects. 
SIMS is a destructive diagnostic, as seen in Fig. \ref{fig:simsbino}. 
We used the SIMS IMS-7f (\textit{CAMECA}), installed at the Institut Jean Lamour (Nancy, France), a double-focusing magnetic sector SIMS that allows detection of isotopes at high mass resolution at the scale of a few microns. We used a high-energy primary ion Cs$^+$ source (impact energy of 15 keV) for efficient sputtering with a beam current of 0.39 nA. The mass spectrometer was tuned to obtain a mass resolving power of $m/\Delta m$ = 3000. We collected and quantified secondary ions of H$^-$, D$^-$, and $^{12}$C$^-$, from the irradiated and non-irradiated tholin spots. 

\begin{figure*}
\centering
\gridline{\fig{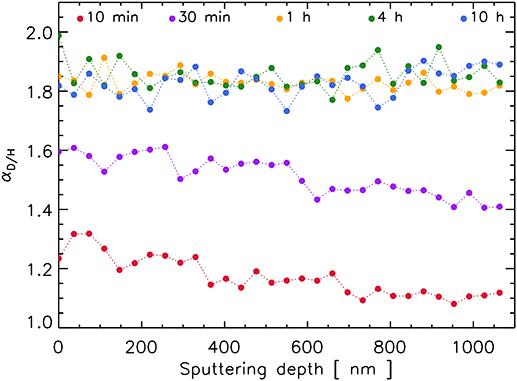}{0.45\textwidth}{(a) $\alpha_{D/H}$ 0.5 keV} 
         \fig{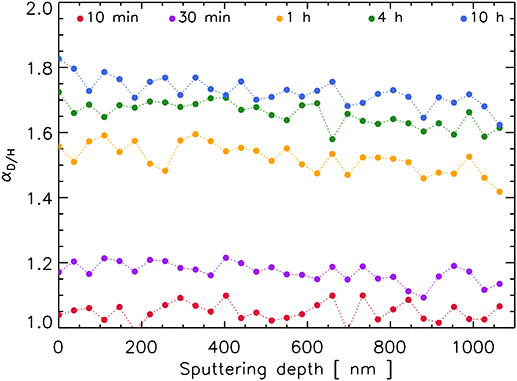}{0.45\textwidth}{(b) $\alpha_{D/H}$ 1.3 keV} 
       }
\gridline{\fig{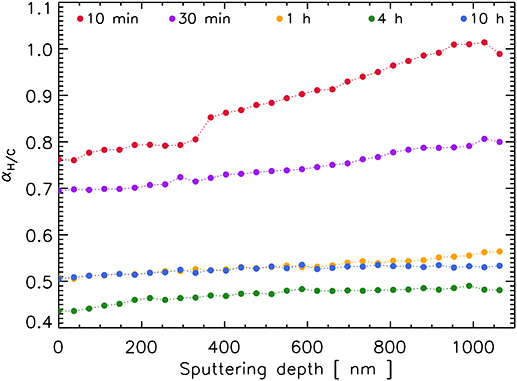}{0.45\textwidth}{(c) $\alpha_{H/C}$ 0.5 keV}
          \fig{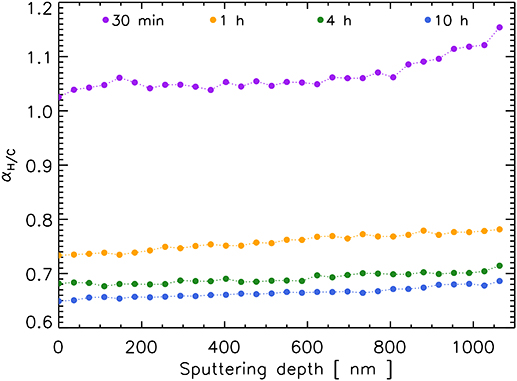}{0.45\textwidth}{(d) $\alpha_{H/C}$ 1.3 keV} 
                    }
\caption{D/H and H/C abundances normalized to the initial tholin ($\alpha_{D/H}$ or $\alpha_{H/C}$) as a function of sputtering depth. For 1.3 keV the $\alpha_{H/C}$ profile after 10 minutes irradiation was discarded due to large systematic errors assigned to localized hydrogenated inhomogeneities on the tholin surface (fractures/wedges), as seen in Fig. \ref{fig:profilo}. \label{fig:sims1}}
\end{figure*}

Before SIMS operation, the irradiated tholin films were gold-coated (thickness $\sim$15 nm). Samples were inserted in the analysis chamber operating at 10$^{-9}$ mbar. The ion beam was rastered over 50 $\times$ 50 $\mu$m$^2$ and, by using an aperture diaphragm, only the inner 35 $\times$ 35 $\mu$m$^2$ were sampled, to reduce the contribution of surface contamination to the H signal. The analyzed area was small enough to accurately probe the 200 $\times$ 200 $\mu$m$^2$ irradiated spots. In between each measurement of irradiated spots, non-irradiated areas on the same tholin film were measured and constitute reference measurements (Fig. \ref{fig:simsbino}). Analyses were saved as depth profiles. Sputtering was stopped as soon as the substrate was reached. The sputtering depth is calibrated using the film thickness obtained via ellipsometry, described in Sec. \ref{sec:ellipso}. The D/H profiles for irradiated spots are reported as isotope fractionation, defined as,
 \begin{equation}
\alpha_{D/H} = \dfrac{D/H_i}{D/H_o}
\end{equation}
where $D/H_o$ is the D/H ratio of the reference tholin measured near the irradiated area and $D/H_i$ is the ratio after a given X-ray dose.  
Similarly, the abundance of H/C for an irradiated spot is parametrized as,
 \begin{equation}
\alpha_{H/C} = \dfrac{H/C_i}{H/C_o}
\end{equation}
where $H/C_o$ corresponds to the reference tholin and $H/C_i$ is the value after a certain irradiation dose. The irradiation dose is defined as,
\begin{equation}
E_{dose} = k \times F,
\end{equation}
where $k$ is the absorption coefficient (Fig. \ref{fig:klp}) and $F$ is the fluence obtained from the total flux integrated in time and divided by the irradiated area.

In Fig. \ref{fig:sims1} (a) we show the $\alpha_{D/H}$ profiles for the 0.5 keV irradiated tholin, which reach a maximum after 1 hour irradiation, corresponding to $\alpha_{D/H}$ = 1.9. Longer exposures do not induce further D-enrichment. We also note that for shorter irradiation times (10 and 30 minutes) the deuterium enrichment decreases linearly in depth. This is evidence to the cumulative dose effect: the regions where the enrichment is largest, hence more affected by irradiation, are found near the surface, while towards the tholin/substrate interface the enrichment decreases and becomes negligible. We hypothesize that the bulk of the tholin is shielding the lower regions from irradiation for short duration experiments. From Fig. \ref{fig:klp}, 0.5 keV photons will have a maximum penetration depth in tholins at $\sim$625 nm. During irradiation, the released secondary electrons (photoelectrons or Auger electrons) will be diffused completely within the tholin volume. 
On the other hand, in Fig. \ref{fig:sims1} (b) we note the linearity of the $\alpha_{D/H}$ profiles for the tholin irradiated at 1.3 keV. At this energy the maximum penetration depth is $>$5 $\mu$m, i.e. for our 1.16 $\mu$m thick samples only $\sim$20 percent of the incident energy will be absorbed.  The penetration length of X-rays in a material depends on the X-ray energy and the absorption index of the material, both affecting the total energy absorbed. For our 1.1 $\mu$m tholin, lower energy X-rays are absorbed more efficiently. This explains the lower fractionation $\alpha_{D/H}$ = 1.7 for 10 h irradiation at 1.3 keV compared to 0.5 keV, and the need for a higher dose to reach the enrichment plateau at 1.3 keV. 

Whereas at 0.5 keV the maximum isotope fractionation is reached after 1 h and does not evolve for longer irradiations, at 1.3 keV isotope fractionation continuously increases from 1 h to 10 h irradiation. In addition, the 10 minute experiment at 1.3 keV induces a negligible D-enrichment. Hence, higher energy X-rays appear less efficient drivers of H-isotope fractionation. 

\begin{figure}[htbp!]
\centering
\epsscale{1.15}
\plotone{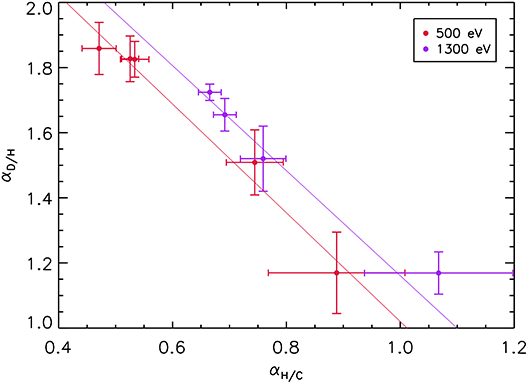}
\caption{Evolution of average normalized D/H vs H/C ratios for X-ray irradiated tholins at 0.5 keV and 1.3 keV. \label{fig:sims4}}
\end{figure}
The $\alpha_{H/C}$ profiles are reported in Fig. \ref{fig:sims1} (c) and (d), and are related to the evolution of the molecular structure during X-ray irradiation. For the shortest irradiation at 0.5 keV, tholins exhibit a modest H-depletion at the surface and a negligible effect near the tholin/substrate interface: $\alpha_{H/C}$ = 0.8 at the surface and increases regularly up to 1 near the interface. Hence, trends for $\alpha_{H/C}$ and $\alpha_{D/H}$ as a function of depth are opposite: D-enrichment is associated to H/C decrease (i.e. loss of H). As the dose increases, dehydrogenation of the organic skeleton increases and becomes more homogeneous across the tholin thickness. At 0.5 keV, $\alpha_{H/C}$ reaches a plateau after 1 hour, while at 1.3 keV the plateau is reached after 4 hours. 

Fig. \ref{fig:sims4} shows the average $\alpha_{D/H}$ versus the average $\alpha_{H/C}$ parameters, for both the 0.5 and 1.3 keV irradiated tholins. These are found from the average $\alpha_{D/H}$ and $\alpha_{H/C}$ values over the entire film thickness in Fig. \ref{fig:sims1}, and the scatter in this value  determines the error bars. The strong correlation between these two parameters confirms that the preferential H over D loss (a reflection of their differences in zero-point energies) during X-ray irradiation is at the origin of the measured deuterium enrichment. 
\begin{figure}[htbp!]
\centering
\epsscale{1.18}
\plotone{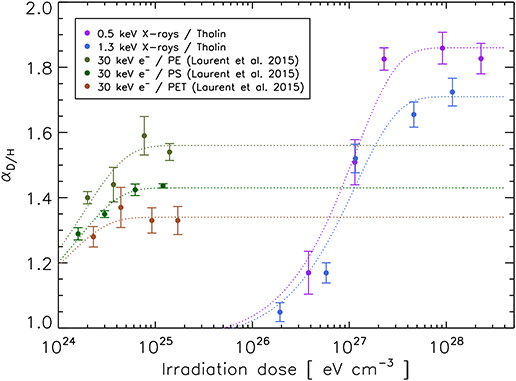}
\caption{Evolution of the normalized D/H abundances as a function of irradiation doses, for 30 keV electrons on organic polymers: polyethylene (PE), polystyrene (PS) and polyethylene terephthalate (PET)   \citep{Laurent2015} and for 0.5 keV and 1.3 keV X-rays on tholins (this work). \label{fig:sims3}}
\end{figure}

Fig. \ref{fig:sims3} shows the evolution of the deuterium enrichment, $\alpha_{D/H}$, as a function of the total X-ray dose for both 0.5 and 1.3 keV. We note that the D-enrichment is slightly larger for 0.5 keV X-rays, which depose a larger energy per volume than 1.3 keV X-rays. 
We compare our results to the evolution observed for 30 keV electron irradiation of organic polymers \citep{Laurent2015}. We first remark that the deposited dose per volume is 1 to 3 orders of magnitude larger in our experiment than for electron irradiation. Secondly, the total maximum D-enrichment (at the plateau)  is about 20\% larger in our study.  X-ray irradiation acts as a bulk process due to the larger penetration depth (compared to electron irradiation) and volume diffusion of secondary electrons, enhancing D-enrichment.

\section{Discussion}

The evolution of tholins induced by X-ray irradiation can be chemically quantified by the sequential depletion and growth of infrared bands, structurally by the analysis of Raman bands, and isotopically by the evolution of the SIMS profiles. For each of these processes, the evolution of the main parameters as a function of the dose can be modeled using first-order rate equations \citep{Laurent2015}, 

\begin{equation}
\dfrac{dy}{dE} = Ae^{-kE_{dose}},
\end{equation}

where $y$ is the chemical, structural, or isotopic parameter, $A$ is a constant, $k$ is the rate constant and $E_{dose}$ is the irradiated dose per volume. We normalized all integrated parameters, such as vibrational band intensities and isotopic abundances, to the starting value (non-irradiated tholin) when possible. The parameters for these fits are listed in Table \ref{tab:fit}. We note that the isotopic rate constant is closely matched by the infrared depletion and the amorphization (Raman broadening) rate constants, pointing at coupled structural and chemical effects of X-ray irradiation, which itself drives the isotopic fractionation, i.e. dehydrogenation of H over D bonds is  accompanied by a change in the organic structure (creation of defects).  

\begin{table}[ht!]
\footnotesize
\renewcommand{\thetable}{\arabic{table}}
\centering
\caption{Fitted infrared, Raman and isotopic evolution parameters as an exponential function of X-ray dose} \label{tab:fit}
\begin{tabular}{cccc}
 \hline
 \hline
     & y$_0$               & A                & k  [ eV$^{-1}$ cm$^3$]                                   \\
 \hline
  C-H     & 0.03($\pm$0.01)  & 0.97($\pm$0.05)  & 1.0($\pm$0.1)$\times$10$^{ -26}$ \\
N-H     & 0.37($\pm$0.05)  & 0.63($\pm$0.04)  & 5.0($\pm$1.8)$\times$10$^{ -27}$ \\
         R-N$\equiv$C & 0.27($\pm$0.04)  & 0.73($\pm$0.06)  & 5.0($\pm$1.6)$\times$10$^{ -27}$ \\
          C-defect     & 1.97($\pm$0.20)  & -0.77($\pm$0.05) & 3.4($\pm$3.0)$\times$10$^{ -27}$ \\
\midrule
 I$_D$/I$_G$   & 0.87($\pm$0.08) & 0.02($\pm$0.01)  & 1.6($\pm$1.3)$\times$10$^{ -28}$ \\
          $\Gamma$-D     & 1.07($\pm$0.08)  & -0.10($\pm$0.01) & 1.3($\pm$0.2)$\times$10$^{ -28}$ \\
         $\Gamma$-G     & 1.07($\pm$0.80)   & -0.09($\pm$0.01) & 3.7($\pm$0.2)$\times$10$^{ -28}$ \\
 \midrule
      $\alpha_{D/H,0.5 keV}$  & 1.86($\pm$0.14)  & -0.90($\pm$0.02)  & 8.8($\pm$1.0)$\times$10$^{ -28}$ \\
          $\alpha_{D/H,1.3 keV}$  & 1.71($\pm$0.21)  & -0.75($\pm$0.01) & 8.7($\pm$0.9)$\times$10$^{ -28}$ \\
          $\alpha_{H/C,0.5 keV}$ & 0.49($\pm$0.04)  & 0.52($\pm$0.01)  & 7.6($\pm$0.7)$\times$10$^{ -28}$ \\
          $\alpha_{H/C,1.3 keV}$  & 0.66($\pm$0.07)  & 0.53($\pm$0.02)  & 7.2($\pm$0.6)$\times$10$^{ -28}$ \\
\bottomrule
\end{tabular}
\end{table}

While thermal degradation \citep{Bonnet2015} and X-ray irradiation of tholins, both induce a similar loss of amines and alkyls (i.e. dehydrogenation of the tholin), there are important differences in their respective evolutions. Raman analysis shows that thermal degradation leads to an increase of polyaromatic units (Fig. \ref{fig:raman3}), while X-irradiation does not affect the average size of polyaromatic units (even if infrared analysis shows that isolated aromatic units may be created). In both cases, defect formation competes with aromatization. For soft X-ray irradiation, the former is the dominant process. 

Disks can be further irradiated by cosmic rays (protons and ions), although their effect may be modulated by winds and magnetic fields in T-Tauri systems \citep{Cleeves2013}. The effect of cosmic rays on organic matter has been studied via laboratory experiments, using low energy or swift heavy ions sources, whose energy can be deposited in solids through nuclear or electronic interactions respectively. \cite{Brunetto2009} examined the effects of 200 - 400 keV ion irradiation on soots showing the destruction of large polyaromatic units. On the other hand, \cite{Costantini2005} reported the carbonization of Kapton films and growth of polyaromatic structures induced by swift heavy ion irradiation. Experiments by \cite{Sabri2015} similarly showed the graphitization of water-covered amorphous carbon induced by 200 keV protons. Together, these experiments point at the strong influence of the starting material structure on its evolution under ionizing radiation. 

Recent disk models by \cite{Cleeves2016} including X-rays and cosmic ray ionization investigated the deuterium enrichment of organic molecules. They argued that since D/H ratios corresponding to the most D-enriched organic materials in the solar system were not attained in the disk, there is a need of some interstellar inheritance. However, these models did not include macromolecular organics. Our experiments show a net loss of heteroatoms from the organic matrix, implying that a fraction of organic solids could be transferred into volatiles (hydrocarbons, radicals, ions). We thus propose that radiolysis and fragmentation of organics can eventually transfer the high D/H ratios of enriched solids back into the gas phase. In addition, as seen by the differences from the irradiation tracks for two X-ray energies (0.5 and 1.3 keV) in Fig. \ref{fig:sims1}, extended studies with tholins of varying structure, using both broadband and hard X-rays sources, will be necessary to fully understand the effect of X-rays on organic matter. 

\section{Astrophysical implications}

\begin{figure}[h!]
\center
\epsscale{1.18}
\plotone{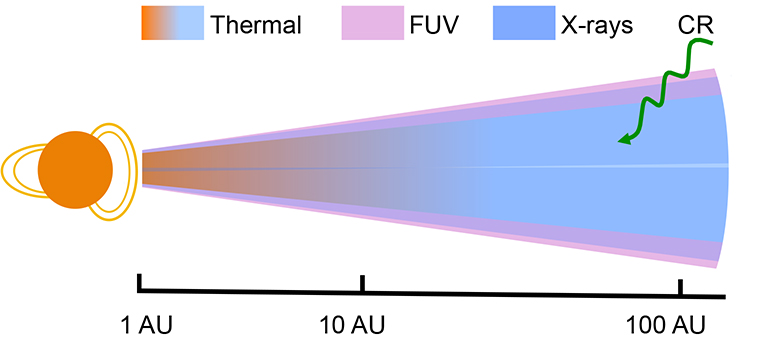}
\caption{Schema of the thermal and non-thermal processes on an XUV illuminated protoplanetary disk. UV photons and X-rays irradiate the disk surface (ionized and molecular layers), but only X-rays reach the mid-plane (regions of ice formation and planetesimal growth) and outer edges. Ionization by cosmic rays (CRs) is represented in green. \label{fig:ppd}}
\end{figure}

Stellar X-ray flux density is an important ionization source in protoplanetary disks, impacting its photochemical evolution. In this experimental study, we have shown that soft X-rays can induce a large  deuterium enrichment in protoplanetary refractory organics. 
We compared the non-thermal effects of X-rays to previous experiments of electron irradiation on polymers \citep{Laurent2015}. X-ray irradiation of tholins induces higher D-enrichment than electron irradiation of organic polymers. This may reflect a better efficiency of X-rays to induce H-D fractionation, or a tendency of tholins to accomodate more defects hence record higher D-enrichment. The efficiency of  irradiation will also depend on the size of the irradiated grains  \citep{Roskosz2016}. Our experiments on $\mu$m thick tholins, comparable to the average size of observed dust grains in protoplanetary disks (from $\mu$m to cm sizes) \citep{Natta2007}, demonstrate that photon-induced isotopic enrichment of organic solids in disks is possible. 

As shown in Fig.~\ref{fig:ppd}, protoplanetary disks undergo irradiation from T-Tauri stellar fields having   strong X-ray and UV components, although the higher energy X-ray photons are able to penetrate more effectively toward the disk mid-plane beyond 10 AU \citep{Nomura2007, Walsh2012}. The X-ray spectra of the classical T Tauri star, TW Hydrae, provides an integrated luminosity from 0.2 to 2 keV of L$_X$ = 2 $\times$10$^{30}$ erg s$^{-1}$ \citep{Kastner2002}. At 10 AU, this amounts to a total exposure of 10$^{26}$ eV cm$^{-2}$ in 1 Myr assuming an X-ray optical depth of $\tau_X$ = 1.  Using synchrotron soft X-rays, we have reached total fluences between 10$^{22}$ - 10$^{24}$ eV cm$^{-2}$, well within the lifetimes of protoplanetary disks. The experimental plateau for D-enrichment is achieved at fluences of $\sim$10$^{24}$ eV cm$^{-2}$ providing an upper limit to D-enrichment of protoplanetary organic matter by soft X-rays.

Compared to UV photons and electron irradiation, soft X-rays and their secondary photoelectrons can strongly modify the bulk of $\mu$m-sized organic particles. In spite of its non-thermal nature, soft X-rays can affect the chemical, structural and isotopic properties of organic macromolecules and solids via photochemical and radiolytical reactions, impacting the cycle of deuterium from dust to volatiles in cold regions of protoplanetary disks.  \\

\vspace{0.05cm}
{ \footnotesize
\textit{Acknowledgments.} X-ray experiments were performed at the SEXTANTS beam line of the SOLEIL synchrotron (project No. 20150772). We thank Ferenc \textsc{Borondics} for access to the IR and Raman spectromicroscopes of the SMIS beam line and Fran\c{c}ois \textsc{Nicolas}
for his help with diagnostics at the Surface Laboratory. N.C. and L.G. thank the European Research Council for funding via the ERC \textit{PrimChem} project (grant agreement No. 636829.)}

\newpage

\begin{table*}[ht!]
\renewcommand{\thetable}{\arabic{table}}
\centering
\caption{Normalized integrated infrared modes ($\alpha_{IR}$) as a function of X-ray dose at 0.5 keV} \label{tab:ir1}
\begin{tabular}{cccccc}
\hline
\hline
         &                  & \multicolumn{4}{c}{Integrated infrared mode [ cm$^{-1}$ ]}                      \\
\hline
time [ s ]& 0.5 keV dose [ eV/cm$^{3}$ ] & C-H                & N-H              & R-N$\equiv$C               & C-defect              \\
\hline
600      & 3.80 $\times$10$^{26}$         & 0.055 ($\pm$0.007) & 0.56 ($\pm$0.04) & 0.45 ($\pm$0.04)  & 1.56 ($\pm$0.15) \\
1800     & 1.14  $\times$10$^{27}$      & 0.036 ($\pm$0.008) & 0.37 ($\pm$0.03) & 0.41 ($\pm$0.03)  & 1.76 ($\pm$0.11)  \\
3600     & 2.28  $\times$10$^{27}$      & 0.042 ($\pm$0.006) & 0.50 ($\pm$0.02) & 0.38 ($\pm$0.03)  & 1.80 ($\pm$0.13) \\
14400    & 9.12  $\times$10$^{27}$          & 0.028 ($\pm$0.008) & 0.32 ($\pm$0.04) & 0.32 ($\pm$0.02)  & 2.09 ($\pm$0.13) \\
36000    & 2.28  $\times$10$^{28}$        & 0.026 ($\pm$0.003) & 0.31 ($\pm$0.05) & 0.30 ($\pm$0.03) & 2.01 ($\pm$0.11) \\
\hline
\end{tabular}
\end{table*}

\begin{table*}[ht!]
\renewcommand{\thetable}{\arabic{table}}
\centering
\caption{Raman parameters as a function of X-ray dose at 0.5 keV. The FWHM of the Raman D and G bands are expressed as $\Gamma$-D and $\Gamma$-G.} \label{tab:raman}
\begin{tabular}{ccccccc}
\hline
\hline
time [ s ] & 0.5 keV dose [ eV/cm$^{3}$ ] & $\omega_D$ [ cm$^{-1}$ ] & $\omega_G$ [ cm$^{-1}$ ] & $\Gamma$-D [ cm$^{-1}$ ] & $\Gamma$-G [ cm$^{-1}$ ] & I$_D$/I$_G$ \\
\hline
600      & 3.80 $\times$10$^{26}$         & 1380.7 ($\pm$ 1.6)     & 1579.3 ($\pm$1.2)     & 263.6 ($\pm$3.2)   & 117.7  ($\pm$2.8) & 0.888 ($\pm$0.010) \\
1800     & 1.14 $\times$10$^{27}$         & 1379.5 ($\pm$ 2.1)     & 1578.8 ($\pm$1.4)     & 258.4  ($\pm$2.9)  & 118.2  ($\pm$2.2) & 0.891 ($\pm$0.012) \\
3600     & 2.28 $\times$10$^{27}$         & 1379.1 ($\pm$1.5)     & 1577.6 ($\pm$1.3)     & 257.4  ($\pm$3.6)  & 121.8  ($\pm$2.1) & 0.889 ($\pm$0.010) \\
14400    & 9.12 $\times$10$^{27}$         & 1379.9 ($\pm$2.5)     & 1571.1 ($\pm$1.2)     & 278.9  ($\pm$3.3)  & 127.5  ($\pm$2.2) & 0.870 ($\pm$0.012) \\
36000    & 2.28 $\times$10$^{28}$        & 1379.1 ($\pm$2.7)     & 1571.0 ($\pm$1.3)     & 279.7  ($\pm$3.1) & 125.1 ($\pm$1.8)  & 0.870 ($\pm$0.010) \\
\hline
\end{tabular}
\end{table*}

\begin{table*}[ht!]
\renewcommand{\thetable}{\arabic{table}}
\centering
\caption{SIMS elemental abundances as a function of X-ray dose at 0.5 keV and 1.3 keV} \label{tab:sims}
\begin{tabular}{cccc|ccc}
\hline
\hline
\multicolumn{1}{l}{} & \multicolumn{3}{c}{0.5 keV}                         & \multicolumn{3}{c}{1.3 keV}                                \\
\hline
time [ s ]             & dose [ eV/cm$^{3}$ ]  & $\alpha_{D/H}$ & $\alpha_{H/C}$ & dose [ eV/cm$^{3}$ ] & $\alpha_{D/H}$    & $\alpha_{H/C}$    \\
\hline
600                  & 3.80 $\times$10$^{26}$      & 1.17  ($\pm$0.12) & 0.89 ($\pm$0.12) & 1.93 $\times$10$^{26}$         & 1.05 ($\pm$0.07)  & NA\textsuperscript{*} \\
1800                 & 1.14 $\times$10$^{27}$  & 1.51 ($\pm$0.10)  & 0.74 ($\pm$0.05) & 5.78 $\times$10$^{26}$          & 1.17 ($\pm$0.10) & 1.07  ($\pm$0.13)  \\
3600                 & 2.28 $\times$10$^{27}$       & 1.83 ($\pm$0.06)  & 0.53 ($\pm$0.03) & 1.16 $\times$10$^{27}$          & 1.52 ($\pm$0.05) & 0.76  ($\pm$0.04)  \\
14400                & 9.12 $\times$10$^{27}$      & 1.86  ($\pm$0.08) & 0.47 ($\pm$0.03) & 4.62 $\times$10$^{27}$         & 1.66 ($\pm$0.05) & 0.69 ($\pm$0.02) \\
36000                & 2.28 $\times$10$^{28}$       & 1.83 ($\pm$0.07)  & 0.53 ($\pm$0.02) & 1.16 $\times$10$^{28}$          & 1.72 ($\pm$0.03) & 0.67 ($\pm$0.02) \\
\hline
\multicolumn{4}{l}{\textsuperscript{*}\footnotesize{Measurement not applicable due to systematic error (fractures in the tholin). }}
\end{tabular}
\end{table*}

\newpage

\bibliography{xtholins}

\begin{thebibliography}{}
\expandafter\ifx\csname natexlab\endcsname\relax\def\natexlab#1{#1}\fi
\providecommand{\url}[1]{\href{#1}{#1}}

\bibitem[{{Albertsson} {et~al.}(2014){Albertsson}, {Semenov}, \&
  {Henning}}]{Albertsson2014}
{Albertsson}, T., {Semenov}, D., \& {Henning}, T. 2014, \apj, 784, 39

\bibitem[{{Al{\'e}on}(2010)}]{Aleon2010}
{Al{\'e}on}, J. 2010, \apj, 722, 1342

\bibitem[{{Allamandola} {et~al.}(1988){Allamandola}, {Sandford}, \&
  {Valero}}]{Allamandola1988}
{Allamandola}, L.~J., {Sandford}, S.~A., \& {Valero}, G.~J. 1988, \icarus, 76,
  225

\bibitem[{{Altwegg} {et~al.}(2015){Altwegg}, {Balsiger}, {Bar-Nun},
  {Berthelier}, {Bieler}, {Bochsler}, {Briois}, {Calmonte}, {Combi}, {De
  Keyser}, {Eberhardt}, {Fiethe}, {Fuselier}, {Gasc}, {Gombosi}, {Hansen},
  {H{\"a}ssig}, {J{\"a}ckel}, {Kopp}, {Korth}, {LeRoy}, {Mall}, {Marty},
  {Mousis}, {Neefs}, {Owen}, {R{\`e}me}, {Rubin}, {S{\'e}mon}, {Tzou}, {Waite},
  \& {Wurz}}]{Altwegg2015}
{Altwegg}, K., {Balsiger}, H., {Bar-Nun}, A., {et~al.} 2015, Science, 347,
  1261952

\bibitem[{{Aresu} {et~al.}(2011){Aresu}, {Kamp}, {Meijerink}, {Woitke}, {Thi},
  \& {Spaans}}]{Aresu2011}
{Aresu}, G., {Kamp}, I., {Meijerink}, R., {et~al.} 2011, \aap, 526, A163

\bibitem[{{Bernard} {et~al.}(2006){Bernard}, {Quirico}, {Brissaud},
  {Montagnac}, {Reynard}, {McMillan}, {Coll}, {Nguyen}, {Raulin}, \&
  {Schmitt}}]{Bernard2006}
{Bernard}, J.-M., {Quirico}, E., {Brissaud}, O., {et~al.} 2006, \icarus, 185,
  301

\bibitem[{{Bonnet} {et~al.}(2015){Bonnet}, {Quirico}, {Buch}, {Thissen},
  {Szopa}, {Carrasco}, {Cernogora}, {Fray}, {Cottin}, {Roy}, {Montagnac},
  {Dartois}, {Brunetto}, {Engrand}, \& {Duprat}}]{Bonnet2015}
{Bonnet}, J.-Y., {Quirico}, E., {Buch}, A., {et~al.} 2015, \icarus, 250, 53

\bibitem[{{Brunetto} {et~al.}(2009){Brunetto}, {Pino}, {Dartois}, {Cao},
  {d'Hendecourt}, {Strazzulla}, \& {Br{\'e}chignac}}]{Brunetto2009}
{Brunetto}, R., {Pino}, T., {Dartois}, E., {et~al.} 2009, \icarus, 200, 323

\bibitem[{{Busemann} {et~al.}(2006){Busemann}, {Young}, {O'D.~Alexander},
  {Hoppe}, {Mukhopadhyay}, \& {Nittler}}]{Busemann2006}
{Busemann}, H., {Young}, A.~F., {O'D.~Alexander}, C.~M., {et~al.} 2006,
  Science, 312, 727

\bibitem[{{Carpentier} {et~al.}(2012){Carpentier}, {F{\'e}raud}, {Dartois},
  {Brunetto}, {Charon}, {Cao}, {d'Hendecourt}, {Br{\'e}chignac}, {Rouzaud}, \&
  {Pino}}]{Carpentier2012}
{Carpentier}, Y., {F{\'e}raud}, G., {Dartois}, E., {et~al.} 2012, \aap, 548,
  A40

\bibitem[{{Carrasco} {et~al.}(2016){Carrasco}, {Jomard}, {Vigneron},
  {Etcheberry}, \& {Cernogora}}]{Carrasco2016}
{Carrasco}, N., {Jomard}, F., {Vigneron}, J., {Etcheberry}, A., \& {Cernogora},
  G. 2016, \planss, 128, 52

\bibitem[{{Ceccarelli} {et~al.}(2014){Ceccarelli}, {Caselli},
  {Bockel{\'e}e-Morvan}, {Mousis}, {Pizzarello}, {Robert}, \&
  {Semenov}}]{Ceccarelli2014}
{Ceccarelli}, C., {Caselli}, P., {Bockel{\'e}e-Morvan}, D., {et~al.} 2014,
  Protostars and Planets VI, 859

\bibitem[{{Ceccarelli} {et~al.}(2005){Ceccarelli}, {Dominik}, {Caux},
  {Lefloch}, \& {Caselli}}]{Ceccarelli2005}
{Ceccarelli}, C., {Dominik}, C., {Caux}, E., {Lefloch}, B., \& {Caselli}, P.
  2005, \apjl, 631, L81

\bibitem[{{Ceccarelli} {et~al.}(2004){Ceccarelli}, {Dominik}, {Lefloch},
  {Caselli}, \& {Caux}}]{Ceccarelli2004}
{Ceccarelli}, C., {Dominik}, C., {Lefloch}, B., {Caselli}, P., \& {Caux}, E.
  2004, \apjl, 607, L51

\bibitem[{{Ceccarelli} {et~al.}(2002){Ceccarelli}, {Vastel}, {Tielens},
  {Castets}, {Boogert}, {Loinard}, \& {Caux}}]{Ceccarelli2002}
{Ceccarelli}, C., {Vastel}, C., {Tielens}, A.~G.~G.~M., {et~al.} 2002, \aap,
  381, L17

\bibitem[{{Ciesla} \& {Sandford}(2012)}]{Ciesla2012}
{Ciesla}, F.~J., \& {Sandford}, S.~A. 2012, Science, 336, 452

\bibitem[{{Cleeves} {et~al.}(2013){Cleeves}, {Adams}, \&
  {Bergin}}]{Cleeves2013}
{Cleeves}, L.~I., {Adams}, F.~C., \& {Bergin}, E.~A. 2013, \apj, 772, 5

\bibitem[{{Cleeves} {et~al.}(2014){Cleeves}, {Bergin}, {Alexander}, {Du},
  {Graninger}, {{\"O}berg}, \& {Harries}}]{Cleeves2014}
{Cleeves}, L.~I., {Bergin}, E.~A., {Alexander}, C.~M.~O.~., {et~al.} 2014,
  Science, 345, 1590

\bibitem[{{Cleeves} {et~al.}(2016){Cleeves}, {Bergin}, {O'D.~Alexander}, {Du},
  {Graninger}, {{\"O}berg}, \& {Harries}}]{Cleeves2016}
{Cleeves}, L.~I., {Bergin}, E.~A., {O'D.~Alexander}, C.~M., {et~al.} 2016,
  \apj, 819, 13

\bibitem[{{Collings} {et~al.}(2004){Collings}, {Anderson}, {Chen}, {Dever},
  {Viti}, {Williams}, \& {McCoustra}}]{Collings2004}
{Collings}, M.~P., {Anderson}, M.~A., {Chen}, R., {et~al.} 2004, \mnras, 354,
  1133

\bibitem[{{Costantini} {et~al.}(2005){Costantini}, {Salvetat}, {Couvreur}, \&
  {Bouffard}}]{Costantini2005}
{Costantini}, J.-M., {Salvetat}, J.-P., {Couvreur}, F., \& {Bouffard}, S. 2005,
  Nuclear Instruments and Methods in Physics Research B, 234, 458

\bibitem[{{Coutens} {et~al.}(2016){Coutens}, {J{\o}rgensen}, {van der Wiel},
  {M{\"u}ller}, {Lykke}, {Bjerkeli}, {Bourke}, {Calcutt}, {Drozdovskaya},
  {Favre}, {Fayolle}, {Garrod}, {Jacobsen}, {Ligterink}, {{\"O}berg},
  {Persson}, {van Dishoeck}, \& {Wampfler}}]{Coutens2016}
{Coutens}, A., {J{\o}rgensen}, J.~K., {van der Wiel}, M.~H.~D., {et~al.} 2016,
  \aap, 590, L6

\bibitem[{{Dartois} {et~al.}(2013){Dartois}, {Engrand}, {Brunetto}, {Duprat},
  {Pino}, {Quirico}, {Remusat}, {Bardin}, {Briani}, {Mostefaoui}, {Morinaud},
  {Crane}, {Szwec}, {Delauche}, {Jamme}, {Sandt}, \& {Dumas}}]{Dartois2013}
{Dartois}, E., {Engrand}, C., {Brunetto}, R., {et~al.} 2013, \icarus, 224, 243

\bibitem[{{Dumas} {et~al.}(2006){Dumas}, {Polack}, {Lagarde}, {Chubar},
  {Giorgetta}, \& {Lefran{\c c}ois}}]{Dumas2006}
{Dumas}, P., {Polack}, F., {Lagarde}, B., {et~al.} 2006, Infrared Physics and
  Technology, 49, 152

\bibitem[{{Duprat} {et~al.}(2010){Duprat}, {Dobric{\u a}}, {Engrand},
  {Al{\'e}on}, {Marrocchi}, {Mostefaoui}, {Meibom}, {Leroux}, {Rouzaud},
  {Gounelle}, \& {Robert}}]{Duprat2010}
{Duprat}, J., {Dobric{\u a}}, E., {Engrand}, C., {et~al.} 2010, Science, 328,
  742

\bibitem[{{Ehrenfreund} {et~al.}(1999){Ehrenfreund}, {Kerkhof}, {Schutte},
  {Boogert}, {Gerakines}, {Dartois}, {D'Hendecourt}, {Tielens}, {van Dishoeck},
  \& {Whittet}}]{Ehrenfreund1999}
{Ehrenfreund}, P., {Kerkhof}, O., {Schutte}, W.~A., {et~al.} 1999, \aap, 350,
  240

\bibitem[{{Feigelson} {et~al.}(2002){Feigelson}, {Garmire}, \&
  {Pravdo}}]{Feigelson2002}
{Feigelson}, E.~D., {Garmire}, G.~P., \& {Pravdo}, S.~H. 2002, \apj, 572, 335

\bibitem[{{Ferrari}(2007)}]{Ferrari2007}
{Ferrari}, A.~C. 2007, Solid State Communications, 143, 47

\bibitem[{{Ferrari} \& {Robertson}(2000)}]{Ferrari2000}
{Ferrari}, A.~C., \& {Robertson}, J. 2000, \prb, 61, 14095

\bibitem[{{Gavilan} {et~al.}(2016){Gavilan}, {J{\"a}ger}, {Simionovici},
  {Lemaire}, {Sabri}, {Foy}, {Yagoubi}, {Henning}, {Salomon}, \&
  {Martinez-Criado}}]{Gavilan2016b}
{Gavilan}, L., {J{\"a}ger}, C., {Simionovici}, A., {et~al.} 2016, \aap, 587,
  A144

\bibitem[{{Glassgold} {et~al.}(1997){Glassgold}, {Najita}, \&
  {Igea}}]{Glassgold1997}
{Glassgold}, A.~E., {Najita}, J., \& {Igea}, J. 1997, \apj, 480, 344

\bibitem[{{Greenberg} {et~al.}(1995){Greenberg}, {Li}, {Mendoza-Gomez},
  {Schutte}, {Gerakines}, \& {de Groot}}]{Greenberg1995}
{Greenberg}, J.~M., {Li}, A., {Mendoza-Gomez}, C.~X., {et~al.} 1995, \apjl,
  455, L177

\bibitem[{{Guilloteau} {et~al.}(2006){Guilloteau}, {Pi{\'e}tu}, {Dutrey}, \&
  {Gu{\'e}lin}}]{Guilloteau2006}
{Guilloteau}, S., {Pi{\'e}tu}, V., {Dutrey}, A., \& {Gu{\'e}lin}, M. 2006,
  \aap, 448, L5

\bibitem[{{Hartogh} {et~al.}(2011){Hartogh}, {Lis}, {Bockel{\'e}e-Morvan}, {de
  Val-Borro}, {Biver}, {K{\"u}ppers}, {Emprechtinger}, {Bergin}, {Crovisier},
  {Rengel}, {Moreno}, {Szutowicz}, \& {Blake}}]{Hartogh2011}
{Hartogh}, P., {Lis}, D.~C., {Bockel{\'e}e-Morvan}, D., {et~al.} 2011, \nat,
  478, 218

\bibitem[{{Henning} \& {Semenov}(2013)}]{Henning2013}
{Henning}, T., \& {Semenov}, D. 2013, Chemical Reviews, 113, 9016

\bibitem[{{Henning} {et~al.}(2010){Henning}, {Semenov}, {Guilloteau}, {Dutrey},
  {Hersant}, {Wakelam}, {Chapillon}, {Launhardt}, {Pi{\'e}tu}, \&
  {Schreyer}}]{Henning2010}
{Henning}, T., {Semenov}, D., {Guilloteau}, S., {et~al.} 2010, \apj, 714, 1511

\bibitem[{{Imanaka} {et~al.}(2004){Imanaka}, {Khare}, {Elsila}, {Bakes},
  {McKay}, {Cruikshank}, {Sugita}, {Matsui}, \& {Zare}}]{Imanaka2004}
{Imanaka}, H., {Khare}, B.~N., {Elsila}, J.~E., {et~al.} 2004, \icarus, 168,
  344

\bibitem[{{Jellison} \& {Modine}(1996)}]{Jellison1996}
{Jellison}, Jr., G.~E., \& {Modine}, F.~A. 1996, Applied Physics Letters, 69,
  371

\bibitem[{{Kastner} {et~al.}(2002){Kastner}, {Huenemoerder}, {Schulz},
  {Canizares}, \& {Weintraub}}]{Kastner2002}
{Kastner}, J.~H., {Huenemoerder}, D.~P., {Schulz}, N.~S., {Canizares}, C.~R.,
  \& {Weintraub}, D.~A. 2002, \apj, 567, 434

\bibitem[{{Kerridge}(1983)}]{Kerridge1983}
{Kerridge}, J.~F. 1983, Earth and Planetary Science Letters, 64, 186

\bibitem[{{Khare} {et~al.}(1984){Khare}, {Sagan}, {Arakawa}, {Suits},
  {Callcott}, \& {Williams}}]{Khare1984}
{Khare}, B.~N., {Sagan}, C., {Arakawa}, E.~T., {et~al.} 1984, \icarus, 60, 127

\bibitem[{{Khare} {et~al.}(1981){Khare}, {Sagan}, {Zumberge}, {Sklarew}, \&
  {Nagy}}]{Khare1981}
{Khare}, B.~N., {Sagan}, C., {Zumberge}, J.~E., {Sklarew}, D.~S., \& {Nagy}, B.
  1981, \icarus, 48, 290

\bibitem[{{Laurent} {et~al.}(2014){Laurent}, {Roskosz}, {Remusat}, {Leroux},
  {Vezin}, \& {Depecker}}]{Laurent2014}
{Laurent}, B., {Roskosz}, M., {Remusat}, L., {et~al.} 2014, \gca, 142, 522

\bibitem[{{Laurent} {et~al.}(2015){Laurent}, {Roskosz}, {Remusat}, {Robert},
  {Leroux}, {Vezin}, {Depecker}, {Nuns}, \& {Lefebvre}}]{Laurent2015}
---. 2015, Nature Communications, 6, 8567

\bibitem[{{Le Guillou} {et~al.}(2013){Le Guillou}, {Remusat}, {Bernard},
  {Brearley}, \& {Leroux}}]{LeGuillou2013}
{Le Guillou}, C., {Remusat}, L., {Bernard}, S., {Brearley}, A.~J., \& {Leroux},
  H. 2013, \icarus, 226, 101

\bibitem[{{Mahjoub} {et~al.}(2012){Mahjoub}, {Carrasco}, {Dahoo}, {Gautier},
  {Szopa}, \& {Cernogora}}]{Mahjoub2012}
{Mahjoub}, A., {Carrasco}, N., {Dahoo}, P.-R., {et~al.} 2012, \icarus, 221, 670

\bibitem[{{Markwick} {et~al.}(2005){Markwick}, {Charnley}, {Butner}, \&
  {Millar}}]{Markwick2005}
{Markwick}, A.~J., {Charnley}, S.~B., {Butner}, H.~M., \& {Millar}, T.~J. 2005,
  \apjl, 627, L117

\bibitem[{{Messenger}(2000)}]{Messenger2000}
{Messenger}, S. 2000, \nat, 404, 968

\bibitem[{{Natta} {et~al.}(2007){Natta}, {Testi}, {Calvet}, {Henning},
  {Waters}, \& {Wilner}}]{Natta2007}
{Natta}, A., {Testi}, L., {Calvet}, N., {et~al.} 2007, Protostars and Planets
  V, 767

\bibitem[{{Nomura} {et~al.}(2007){Nomura}, {Aikawa}, {Tsujimoto}, {Nakagawa},
  \& {Millar}}]{Nomura2007}
{Nomura}, H., {Aikawa}, Y., {Tsujimoto}, M., {Nakagawa}, Y., \& {Millar}, T.~J.
  2007, \apj, 661, 334

\bibitem[{{{\"O}berg} {et~al.}(2012){{\"O}berg}, {Qi}, {Wilner}, \&
  {Hogerheijde}}]{Oberg2012}
{{\"O}berg}, K.~I., {Qi}, C., {Wilner}, D.~J., \& {Hogerheijde}, M.~R. 2012,
  \apj, 749, 162

\bibitem[{{Parise} {et~al.}(2011){Parise}, {Belloche}, {Du}, {G{\"u}sten}, \&
  {Menten}}]{Parise2011}
{Parise}, B., {Belloche}, A., {Du}, F., {G{\"u}sten}, R., \& {Menten}, K.~M.
  2011, \aap, 526, A31

\bibitem[{{Parise} {et~al.}(2006){Parise}, {Ceccarelli}, {Tielens}, {Castets},
  {Caux}, {Lefloch}, \& {Maret}}]{Parise2006}
{Parise}, B., {Ceccarelli}, C., {Tielens}, A.~G.~G.~M., {et~al.} 2006, \aap,
  453, 949

\bibitem[{{Preibisch} {et~al.}(2005){Preibisch}, {Kim}, {Favata}, {Feigelson},
  {Flaccomio}, {Getman}, {Micela}, {Sciortino}, {Stassun}, {Stelzer}, \&
  {Zinnecker}}]{Preibisch2005}
{Preibisch}, T., {Kim}, Y.-C., {Favata}, F., {et~al.} 2005, \apjs, 160, 401

\bibitem[{{Qi} {et~al.}(2008){Qi}, {Wilner}, {Aikawa}, {Blake}, \&
  {Hogerheijde}}]{Qi2008}
{Qi}, C., {Wilner}, D.~J., {Aikawa}, Y., {Blake}, G.~A., \& {Hogerheijde},
  M.~R. 2008, \apj, 681, 1396

\bibitem[{{Remusat} {et~al.}(2006){Remusat}, {Palhol}, {Robert}, {Derenne}, \&
  {France-Lanord}}]{Remusat2006}
{Remusat}, L., {Palhol}, F., {Robert}, F., {Derenne}, S., \& {France-Lanord},
  C. 2006, Earth and Planetary Science Letters, 243, 15

\bibitem[{{Remusat} {et~al.}(2009){Remusat}, {Robert}, {Meibom}, {Mostefaoui},
  {Delpoux}, {Binet}, {Gourier}, \& {Derenne}}]{Remusat2009}
{Remusat}, L., {Robert}, F., {Meibom}, A., {et~al.} 2009, \apj, 698, 2087

\bibitem[{{Robert} \& {Epstein}(1982)}]{Robert1982}
{Robert}, F., \& {Epstein}, S. 1982, \gca, 46, 81

\bibitem[{{Roberts} {et~al.}(2002){Roberts}, {Fuller}, {Millar}, {Hatchell}, \&
  {Buckle}}]{Roberts2002}
{Roberts}, H., {Fuller}, G.~A., {Millar}, T.~J., {Hatchell}, J., \& {Buckle},
  J.~V. 2002, \aap, 381, 1026

\bibitem[{{Roskosz} {et~al.}(2016){Roskosz}, {Laurent}, {Leroux}, \&
  {Remusat}}]{Roskosz2016}
{Roskosz}, M., {Laurent}, B., {Leroux}, H., \& {Remusat}, L. 2016, The
  Astrophysical Journal, 832, 55

\bibitem[{{Roueff} {et~al.}(2015){Roueff}, {Loison}, \& {Hickson}}]{Roueff2015}
{Roueff}, E., {Loison}, J.~C., \& {Hickson}, K.~M. 2015, \aap, 576, A99

\bibitem[{{Sabri} {et~al.}(2015){Sabri}, {Baratta}, {J{\"a}ger}, {Palumbo},
  {Henning}, {Strazzulla}, \& {Wendler}}]{Sabri2015}
{Sabri}, T., {Baratta}, G.~A., {J{\"a}ger}, C., {et~al.} 2015, \aap, 575, A76

\bibitem[{Sacchi {et~al.}(2013)Sacchi, Jaouen, Popescu, Gaudemer, Tonnerre,
  Chiuzbaian, Hague, Delmotte, Dubuisson, Cauchon, Lagarde, \&
  Polack}]{Sacchi2013}
Sacchi, M., Jaouen, N., Popescu, H., {et~al.} 2013, Journal of Physics:
  Conference Series, 425, 072018

\bibitem[{{Sagan} \& {Khare}(1979)}]{Sagan1979}
{Sagan}, C., \& {Khare}, B.~N. 1979, \nat, 277, 102

\bibitem[{{Sagan} {et~al.}(1993){Sagan}, {Khare}, {Thompson}, {McDonald},
  {Wing}, {Bada}, {Vo-Dinh}, \& {Arakawa}}]{Sagan1993}
{Sagan}, C., {Khare}, B.~N., {Thompson}, W.~R., {et~al.} 1993, \apj, 414, 399

\bibitem[{{Sandford} {et~al.}(2001){Sandford}, {Bernstein}, \&
  {Dworkin}}]{Sandford2001}
{Sandford}, S.~A., {Bernstein}, M.~P., \& {Dworkin}, J.~P. 2001, Meteoritics
  and Planetary Science, 36, 1117

\bibitem[{{Sciamma-O'Brien} {et~al.}(2012){Sciamma-O'Brien}, {Dahoo},
  {Hadamcik}, {Carrasco}, {Quirico}, {Szopa}, \& {Cernogora}}]{Sciamma2012}
{Sciamma-O'Brien}, E., {Dahoo}, P.-R., {Hadamcik}, E., {et~al.} 2012, \icarus,
  218, 356

\bibitem[{{Szopa} {et~al.}(2006){Szopa}, {Cernogora}, {Boufendi}, {Correia}, \&
  {Coll}}]{Szopa2006}
{Szopa}, C., {Cernogora}, G., {Boufendi}, L., {Correia}, J.~J., \& {Coll}, P.
  2006, \planss, 54, 394

\bibitem[{T\"{o}rm\"{a} {et~al.}(2014)T\"{o}rm\"{a}, Kostamo, Sipil\"{a},
  Mattila, Kostamo, Kostamo, Lipsanen, Laubis, Scholze, Nelms, Shortt, \&
  Bavdaz}]{Torma2014}
T\"{o}rm\"{a}, P.~T., Kostamo, J., Sipil\"{a}, H., {et~al.} 2014, IEEE
  Transactions on Nuclear Science, 61, 695

\bibitem[{{van der Tak} {et~al.}(2009){van der Tak}, {M{\"u}ller}, {Harding},
  \& {Gauss}}]{VanderTak2009}
{van der Tak}, F.~F.~S., {M{\"u}ller}, H.~S.~P., {Harding}, M.~E., \& {Gauss},
  J. 2009, \aap, 507, 347

\bibitem[{{van Dishoeck} {et~al.}(2003){van Dishoeck}, {Thi}, \& {van
  Zadelhoff}}]{VanDishoeck2003}
{van Dishoeck}, E.~F., {Thi}, W.-F., \& {van Zadelhoff}, G.-J. 2003, \aap, 400,
  L1

\bibitem[{{Walsh} {et~al.}(2012){Walsh}, {Nomura}, {Millar}, \&
  {Aikawa}}]{Walsh2012}
{Walsh}, C., {Nomura}, H., {Millar}, T.~J., \& {Aikawa}, Y. 2012, \apj, 747,
  114

\bibitem[{{Williams} \& {Cieza}(2011)}]{Williams2011}
{Williams}, J.~P., \& {Cieza}, L.~A. 2011, \araa, 49, 67

\bibitem[{{Yang} \& {Epstein}(1983)}]{Yang1983}
{Yang}, J., \& {Epstein}, S. 1983, \gca, 47, 2199

\end{thebibliography}

\end{document}